\documentclass{INTERSPEECH2023}
\usepackage{url}
\usepackage{hyperref} 
\usepackage{diagbox}
\interspeechcameraready 
\title{HierVST: Hierarchical Adaptive Zero-shot Voice Style Transfer}

\name{Sang-Hoon Lee$^{*}$\thanks{$^{*}$Equal contribution}, Ha-Yeong Choi$^{*}$, Hyung-Seok Oh, Seong-Whan Lee$^{\dagger}$\thanks{$^\dagger$Corresponding author}}
\address{
Department of Artificial Intelligence, Korea University, Seoul, Korea
}
\email{\{sh\_lee, hayeong, hs\_oh, sw.lee\}@korea.ac.kr}

\begin{document}

\maketitle
 
\begin{abstract}
Despite rapid progress in the voice style transfer (VST) field, recent zero-shot VST systems still lack the ability to transfer the voice style of a novel speaker. In this paper, we present HierVST, a hierarchical adaptive end-to-end zero-shot VST model. Without any text transcripts, we only use the speech dataset to train the model by utilizing hierarchical variational inference and self-supervised representation. In addition, we adopt a hierarchical adaptive generator that generates the pitch representation and waveform audio sequentially. Moreover, we utilize unconditional generation to improve the speaker-relative acoustic capacity in the acoustic representation. With a hierarchical adaptive structure, the model can adapt to a novel voice style and convert speech progressively. The experimental results demonstrate that our method outperforms other VST models in zero-shot VST scenarios. Audio samples are available at \url{https://hiervst.github.io/}.

\end{abstract}

\section{Introduction}
Recently, voice conversion (VC) systems \cite{qian2019autovc, yang22f_interspeech,lee2022duration,bilinski22_interspeech,choi2023dddm} have shown rapid progress, with significant performance in 
 voice style transfer (VST). Concurrently, progress in neural vocoder models \cite{kong2020hifi, kim21f_interspeech, jang21_interspeech, koizumi22_interspeech, yoneyama2023source} has accelerated the development of VC systems because of their ability to generate high-fidelity waveform audio, and the end-to-end VC systems \cite{polyak21_interspeech, kim2021conditional,liu22c_interspeech,choi2023nansy} have garnered significant interest by generating high-quality converted waveform audio by combining the VC and neural vocoder. However, end-to-end models still have low speaker adaptation performance and require text transcripts to disentangle linguistic representations from speech. Hence, there is a limitation where a paired text-audio dataset is required.

 To utilize a non-parallel speech dataset, AutoVC \cite{qian2019autovc} introduces an information bottleneck on content representation to disentangle the content and style information, and train the model with only self-reconstruction loss. However, there is a trade-off between audio quality and VST performance according to the information bottleneck size and there is a difficulty in choosing the appropriate information bottleneck size. F0-AutoVC \cite{qian2020f0} extends AutoVC to use an additional pitch contour from the source speech, and transforms the normalized pitch contour to the target pitch contour using the target speech statistics. Despite these pitch contour guides, most F0 extraction algorithms have a problem of extracting inaccurate F0 causing unnaturalness by generating a noisy sound and a voice style different from target speaker. 

 \cite{polyak21_interspeech,maimon2022speaking} utilizes a discrete unit of self-supervised speech representation and quantized representation of normalized F0 to reconstruct speech, and convert the speech only by replacing the speaker representation. NANSY \cite{choi2021neural} utilizes continuous self-supervised speech representation, and introduces a speech perturbation to acquire only the linguistic representation from speech. HierSpeech \cite{lee2022hierspeech} also uses self-supervised speech representation to extract the linguistic representation from speech, but text transcripts are required to regularize the linguistic representation to contain only linguistic information. Diffusion-based VC systems \cite{popov2022diffusionbased, sadekova22_interspeech} also show an improvement in generative performance. However, they also require text transcripts to train the average-Mel encoder from the extracted phoneme alignment \cite{mcauliffe2017montreal}. In addition, most models still have limitations in zero-shot VC, resulting from a lack of ability in VST.

 To address the above problems, we propose HierVST, a hierarchical adaptive end-to-end VST system. We adopt a multi-path self-supervised speech representation from a single speech by restoring the speaker-agnostic linguistic representation from perturbed speech and extracting the speaker-related linguistic representation from the original speech. We also introduce a hierarchical adaptive generator (HAG) with source modeling, and connect multiple representations through hierarchical variational inference. We found that hierarchical adaptation is the key to the success of zero-shot VC. Moreover, we present prosody distillation for enhanced linguistic representation and unconditional generation on the HAG to improve the acoustic capacity on acoustic representation for better speaker adaptation. The experimental results demonstrate that our model outperforms the others in terms of audio quality and speaker similarity on the zero-shot VST without any text transcripts. 

\begin{figure*}[!t]
  \centering
\centerline{\includegraphics[width=0.93\textwidth]{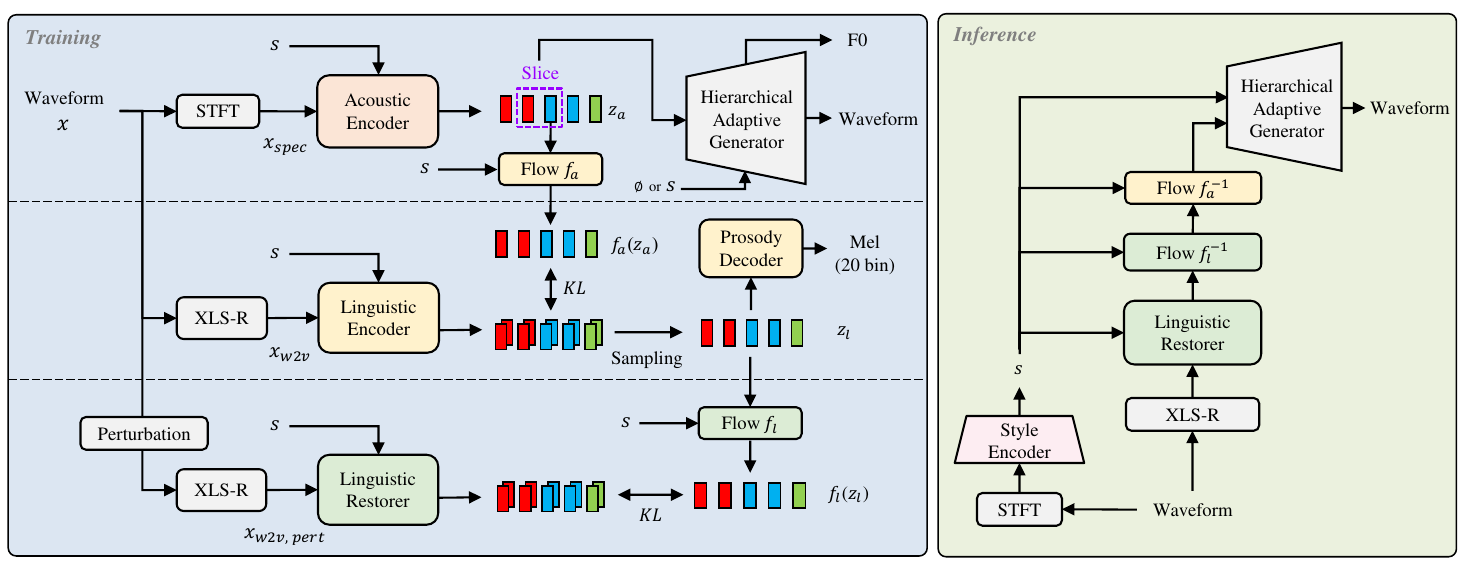}}\vspace{-0.3cm}
\caption{Overall framework of HierVST}
\label{framework}\vspace{-0.4cm}
\end{figure*}

\section{HierVST}
We present a hierarchical adaptive end-to-end VST system, HierVST. For untranscribed voice conversion, we introduce a multi-path self-supervised speech representation, and adopt hierarchical variational inference to connect the speech representations. Furthermore, we introduce a HAG, prosody distillation, and unconditional generation for better speaker adaptation. The details are described in the following subsections.

\subsection{Speech representation}
For voice conversion, we first decompose the speech into perturbed linguistic representation, linguistic representation, and acoustic representation and resynthesize the speech from disentangled representations. Following \cite{kim2021conditional,lee2022hierspeech}, we use a high-resolution linear spectrogram to extract the acoustic representation. For speaker adaptation, we also extract the style representation from the Mel-spectrogram.

\subsubsection{Linguistic representation}
Following \cite{lee2022hierspeech}, the wav2vec feature $x_{w2v}$ is extracted from the representation from the middle layer of XLS-R, which is a pre-trained self-supervised model with a large-scale cross-lingual speech dataset. For speech disentanglement, we introduce a multi-path self-supervised speech representation by utilizing data perturbation \cite{choi2021neural} to reduce the content-irrelevant features from the same self-supervised speech model. The extracted $x_{w2v,pert}$ from the perturbed speech is fed to the linguistic restorer to restore the linguistic representation. The extracted $x_{w2v}$ is fed to the linguistic encoder to extract an enhanced linguistic representation. 

\subsubsection{Style representation}
\label{stylerepresentation}
For global voice style representation (timbre information), we extract the style representation from the Mel-spectrogram. The style encoder \cite{min2021meta} is utilized to extract the style representation which is an averaged style vector of the single sentence, and this encoder is jointly trained with the model in an end-to-end manner. For hierarchical style adaptation, this style representation is fed to all networks including the linguistic restorer, linguistic encoder, acoustic encoder, normalizing flow modules, and HAG. For the zero-shot VST scenario, we do not use the speaker ID information, and we only extract the style representation from the speech.

\subsection{Hierarchical variational autoencoder}
We adopt the structure of HierSpeech \cite{lee2022hierspeech} for an end-to-end VST system replacing the text encoder with a linguistic restorer. We utilize the perturbed linguistic representation $x_{w2v,pert}$ as conditional information $c$ to hierarchically generate waveform audio. We additionally use the enhanced linguistic representation from the self-supervised representation of the original waveform, which is not perturbed. Moreover, the raw waveform audio is reconstructed from the acoustic representation which is extracted using a linear spectrogram during the training. To connect acoustic and multi-path linguistic representations, we utilize hierarchical variation inference, and the optimization objective of HierVST can be defined as follows:
\begin{equation}
\begin{split}
    \log{p_{\theta}}(x|c)&\geq \mathbb{E}_{q_{\phi}(z|x)}\Big[\log p_{\theta_d}(x|z_a)\\&-\log \frac{q_{\phi_{a}}(z_a|x_{spec})}{p_{\theta_a}(z_a|z_{l})} -\log \frac{q_{\phi_{l}}(z_l|x_{w2v})}{p_{\theta_l}(z_l|c)} \Big]
    \end{split}
\end{equation}
where $q_{\phi_a}(z_a|x_{spec})$ and $q_{\phi_l}(z_l|x_{w2v})$ are the approximate posteriors for the acoustic and linguistic representations respectively. $p_{\theta_l}(z_l|c)$ represents a prior distribution of linguistic latent variables $z_l$ given condition $c$, $p_{\theta_a}(z_a|z_{l})$ denotes a prior distribution on acoustic latent variables $z_a$, and $p_{\theta_d}(x|z_a)$ is the likelihood function represented by a HAG that produces data $x$ given latent variables $z_a$. In addition, we use the normalizing flow to improve the expressiveness of each linguistic representation. For the reconstruction loss, we use multiple reconstruction terms of a HAG as described in the following subsection.

\subsection{Hierarchical adaptive generator}
\begin{figure}[!t]
  \centering
\centerline{\includegraphics[width=0.96\columnwidth]{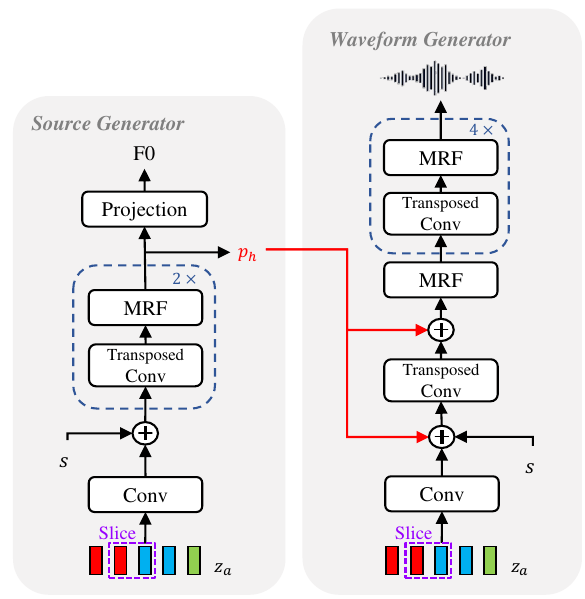}}
\vspace{-0.2cm}
\caption{Hierarchical adaptive generator}
\vspace{-0.7cm}\label{HAG}
\end{figure}

For end-to-end VC, we additionally introduce the HAG $G$ which consists of the source generator $G_s$ and waveform generator $G_w$ as illustrated in Figure \ref{HAG}. The generated representations including acoustic representation $z_a$, style representation $s$ are fed to $G_s$, and $G_s$ generates the refined pitch representation $p_h$ and auxiliary F0 predictor is used to enforce the F0 information on $p_h$ as follows:

\begin{equation}
\label{Pitch_loss}
   L_{pitch} = \lVert p_x-G_s(z_a, s)\rVert_1,
\end{equation}
where $p_x$ is the ground-truth (GT) log-scale F0. Subsequently, $G_w$ synthesizes the waveform audio from $z_a, p_h, s$ hierarchically, and we use the reconstruction loss between the GT and generated Mel-spectrogram transformed from waveform audio using STFT with Mel-filter $\psi$ as follows:
\begin{equation}
\label{Mel_loss}
   L_{STFT} = \lVert \psi(x)-\psi(G_w(z_a, p_h, s))\rVert_1.
\end{equation}

In addition, we utilize adversarial training \cite{lee2021multi,chung21_interspeech} to improve audio quality. We adopt the multi-period discriminator (MPD) \cite{kong2020hifi}\footnote{When we remove MPD for fast training, we observed that audio quality perceptually decreases.} and the multi-scale STFT discriminator (MS-STFTD) \cite{defossez2022high} which can reflect the characteristic of real and imaginary components from a complex-valued STFT as: 
\begin{equation}
  \mathcal{L}_{adv}(D) = \mathbb{E}_{(x,z_a)}\Big[(D(x)-1)^2 + D(G(z_a, s))^2 \Big],
\end{equation}
\begin{equation}
  \mathcal{L}_{adv}(\phi_a, \theta_d) = \mathbb{E}_{(z_a)}\Big[(D(G(z_a, s))-1)^2 \Big]
\end{equation}
\subsection{Prosody distillation}
We introduce prosody distillation to extract the enhanced linguistic representation $z_l$ from the linguistic encoder. $z_l$ is fed to the prosody decoder which generates the first 20 bins of Mel-spectrogram containing the prosody representation. Unlike ProsoSpeech \cite{ren2022prosospeech} which restricts the speaker information from the prosody vector, we make $z_l$ acquire speaker-related prosody information for enhanced linguistic information. We use the prosody loss $\mathcal{L}_{prosody}$ which minimizes the $l1$ distance between the 20 bin of GT and reconstructed Mel-spectrogram.

\subsection{Unconditional generation}
For speaker adaptation, we use style representation as a condition for the network within the entire framework as mentioned in Section \ref{stylerepresentation}. We observed that the speaker adaptation is performed mainly in the HAG. Hence, we introduce an unconditional generation in the hierarchical generator to increase the speaker characteristic on the acoustic representation for progressive speaker adaptation. Following \cite{kim2022guided2}, we simply replace the style representation $s$ with the null speaker embedding $\varnothing$ by a 10\% chance, so we can treat the model as a conditional and unconditional model in the single model.                             

\section{Experiment and result}
\subsection{Dataset and preprocessing}
We use the large-scale multi-speaker dataset, LibriTTS \cite{zen2019libritts} to train the model ($\textit{train-clean-360}$ and $\textit{train-clean-100}$), which consists of about 300-hours of speech for 1,151 speakers. We use \textit{dev-clean} subset for validation. To evaluate the zero-shot VST task, we utilize the VCTK dataset \cite{veaux2017superseded}. For both datasets, we downsample audio to 16 kHz. For self-supervised speech representation, the downsampled audio is fed to the XLS-R model to extract the linguistics-related representation from the middle layer of XLS-R, and this representation is a sequence of 1024-dimensional vectors downsampled from 16 kHz audio (320$\times$ downsampled scale). We also utilize high-resolution F0 which is a sequence of F0 extracted from audio (80$\times$ downsampled scale). For the Mel-spectrogram, we transform audio using the short-time Fourier transform (STFT) with a hop size of 320, a window size of 1,280, an FFT size of 1,280, and 80 bins of Mel-filter.

\subsection{Training}
\vspace{-0.1cm}We use the AdamW optimizer \cite{loshchilov2018decoupled} with the same setting of \cite{lee2022hierspeech}. We train HierVST with a batch size of 128 for 600k steps on four NVIDIA A100 GPUs (six days). For one-shot VST, we fine-tune the model with only a single sample of novel speakers for 1,000 steps and we initialize the same AdamW optimizer but a lower learning rate of $1\times10^{-4}$. We train the model for ablation study with a batch size of 64 on two A100 GPUs for 300k steps. For efficient training, we use a segment audio of 61,440 frames for input audio and utilize the windowed generator training with additional sliced audio of 9,600 frames.   
 
\vspace{-0.1cm}\subsection{Implementation details}
\vspace{-0.1cm}The linguistic restorer, linguistic encoder, and acoustic encoder consist of 16 layers of non-causal WaveNet with 192 hidden dimensions. The flow modules including $f_l$ and $f_a$ consist of four affine coupling layers with four layers of WaveNet. 
 For the HAG, the source generator consists of two upsampling layers of [2,2] and two multi-receptive field fusion (MRF) blocks, and the waveform generator consists of HiFi-GAN \cite{kong2020hifi} and a conditional layer from the representation of the source generator. We use the upsampling rate of [4,5,4,2,2] and an initial channel of 512. For the discriminator, we use the MPD \cite{kong2020hifi} and the MS-STFTD \cite{defossez2022high} with five different sizes of window([2048,1024,512,256,128]). We use a shallow feed-forward transformer network with two layers and 768 hidden dimensions for prosody distillation. For the unconditional generation, we set the ratio of unconditional generation $p_{uncond}$ to 0.1. We fine-tune the model only with the conditional generation. The number of entire model parameter for inference is 45M. 

 \begin{table}[t]
  \centering
    \caption{Many-to-many VST results from LibriTTS dataset} \vspace{-0.3cm}
  \label{table1}
      \resizebox{1\columnwidth}{!}{
  \begin{tabular}{l|cc|cc|cc}
    \toprule
     Method &  nMOS & sMOS  & CER  & WER & EER &SECS\\
    \midrule
     GT  &  4.55$\pm0.04$ & 3.97$\pm0.01$ & 0.54 & 1.84 & - &- \\
    HiFi-GAN \cite{kong2020hifi}&  4.17$\pm0.04$ & 3.86$\pm0.03$ & 0.60 & 2.19 & - &0.986 \\
    \midrule
    AutoVC \cite{qian2019autovc} & 2.57$\pm0.06$ & 2.21$\pm0.05$ & 5.34 & 8.53 & 33.30 &0.703\\

      VoiceMixer \cite{lee2021voicemixer} & 2.84$\pm0.06$ & 2.49$\pm0.05$ & 2.39 & 4.20 & 16.00 &0.779   \\
    DiffVC \cite{popov2022diffusionbased} &  3.50$\pm0.06$ & 3.02$\pm0.05$  & 7.99 & 13.92 & 11.00 &0.817 \\

         \midrule
     SR \cite{polyak21_interspeech} &  2.75$\pm0.06$ & 2.32$\pm0.05$  & 6.63 & 11.72 & 33.30 &0.693\\
     YourTTS  \cite{casanova2022yourtts}  & 2.83$\pm0.06$ &  2.35$\pm0.04$ & 5.43 & 8.79 & 8.00& 0.769 \\
     HierVST (Ours) & \textbf{4.06$\pm$0.05} & \textbf{3.29$\pm$0.04} &\textbf{0.84}& \textbf{2.22}& \textbf{5.25}& \textbf{0.827}\\
    \bottomrule
  \end{tabular}
} \vspace{-0.7cm}
\end{table}

\vspace{-0.1cm}\subsection{Many-to-many VST}
\vspace{-0.1cm}We compared our model with five baseline models: (1) AutoVC \cite{qian2019autovc}, information bottleneck based VC model. (2) VoiceMixer \cite{lee2021voicemixer}, similarity-based information bottleneck and adversarial training based VC model. (3) DiffVC \cite{popov2022diffusionbased}, diffusion-based VC model. (4) Speech Resynthesis (SR) \cite{polyak21_interspeech}, an end-to-end model using discrete speech units. (5) YourTTS\footnote{We used an official pre-trained model. However, this model was trained with LibriTTS, VCTK, and an additional dataset. In addition, YourTTS utilizes text transcripts for training.} \cite{casanova2022yourtts}, an end-to-end speech synthesis model, based on VITS \cite{kim2021conditional}. 
% 5) HierSpeech-U \cite{lee2022hierspeech}, an end-to-end speech synthesis model\footnote{The public-available VC model, FreeVC \cite{lifreevc}, has a same architecture of HierSpeech-U without utilizing the WavLM instead of XLS-R. However, when using the data perturbation of \cite{choi2021neural}, the WavLM representation can not reconstruct the speech with CER of 58\%.}
% We train all models with the same dataset. We train the HierSpeech-U with data perturbation as an flow-based end-to-end VC model which uses the same input representation of HierVST.
Following \cite{lee2022hierspeech}, we conduct naturalness mean opinion score (nMOS) and similarity MOS (sMOS) for subjective evaluation metrics. To evaluate linguistic consistency, we also calculate the character error rate (CER) and word error rate (WER) by Whisper large model \cite{radford2022robust}. For the speaker similarity measurements, we calculate the equal error rate (EER) of the automatic speech recognition model \cite{kwon2021ins} and speaker embedding cosine similarity (SECS) of Resemblyzer\footnote{https://github.com/resemble-ai/Resemblyzer} between the target and converted speech.

Table \ref{table1} shows that our model achieves a significant improvement in all evaluation metrics. Specifically, audio quality improved and the speaker adaptation quality increased in terms of nMOS and sMOS, respectively. Also, our model can convert the speech with a small loss of content information, where the CER and WER are much lower than others even though our model is trained without text transcripts. The objective metrics for speaker similarity also show that our model can adapt well to target voice style. Although HierVST has a similar structure using variational inference augmented with the normalizing flow \cite{casanova2022yourtts}, our hierarchical structure has better speaker adaptation and audio quality including naturalness and pronunciation.  

\begin{table}[t]
  \centering
      \caption{Zero-shot VST results on unseen speakers from VCTK}
  \label{zvct2}\vspace{-0.3cm}
      \resizebox{0.92\columnwidth}{!}{
  \begin{tabular}{l|cc|cc|cc}
    \toprule
     Method &  nMOS  & sMOS  & CER  & WER  & EER &SECS  \\
    \midrule
     GT  & 4.42$\pm0.04$ & 3.98$\pm0.01$ & 0.21 & 2.17 & - & - \\
     HiFi-GAN \cite{kong2020hifi} & 4.15$\pm0.05$ & 3.91$\pm0.02$ & 0.21 & 2.17 & - & 0.989 \\
     \midrule
      AutoVC \cite{qian2019autovc} & 2.47$\pm0.05$ & 1.79$\pm0.05$  &5.14 & 10.55& 37.32 &0.715  \\
      VoiceMixer \cite{lee2021voicemixer} &  2.79$\pm0.05$ & 2.28$\pm0.06$ & \textbf{1.08} & \textbf{3.31} & 20.75&0.797  \\
      DiffVC \cite{popov2022diffusionbased} & 3.51$\pm0.07$ & 2.44$\pm0.05$ & 6.92 & 13.19 & 24.01 & 0.785  \\
    \midrule
      SR \cite{polyak21_interspeech} & 2.27$\pm0.05$ & 2.15$\pm0.06$ & 2.12 & 6.18 & 27.24 &0.750 \\
   YourTTS \cite{casanova2022yourtts}  & 2.69$\pm0.05$ & 2.31$\pm0.06$ & 2.42 & 6.08 & \textbf{4.02} & 0.848 \\
    HierVST (Ours) & \textbf{4.12$\pm$0.05} & \textbf{2.70$\pm$0.06} &1.14&3.46& \textbf{5.06}& \textbf{0.850} \\
    \bottomrule
  \end{tabular}
   }\vspace{-0.5cm}
\end{table}
\vspace{-0.1cm}\subsection{Zero-shot VST}
\vspace{-0.1cm}We compared the performance of zero-shot VST on the VCTK dataset. Table \ref{zvct2} shows that only our model can adapt to novel speakers in terms of EER and SECS. Note that YourTTS is trained with the VCTK dataset so the VST scenario of YourTTS is not the zero-shot VST. Nonetheless, the zero-shot speaker adaptation results of our model show a speaker adaptation quality similar to that of YourTTS in terms of EER and SECS. Furthermore, our model also achieves much better performance on both subjective metrics than others, and this means our model robustly converts speech even in the zero-shot VST scenario with a hierarchical adaptive structure. 

\begin{table}[h]\vspace{-0.25cm}
\caption{One-shot VST results on VCTK dataset according to the number of fine-tuning steps}\vspace{-0.3cm}
  \label{oneshotVST}
  \centering
      \resizebox{0.94\columnwidth}{!}{
  \begin{tabular}{c|c|ccccc}
    \toprule
    \diagbox{Metric}{Step} & 0 (zero-shot) & 100  & 300  & 500 & 1000 & 1500 \\
    \midrule
    CER ($\downarrow$)  & 1.14  & 0.74    & 0.76    & $\textbf{0.66}$ & 0.79 & 1.13 \\
    WER ($\downarrow$)  &  3.46  & 2.77    & 2.85    & $\textbf{2.72}$ & 3.05 & 3.63  \\
    EER ($\downarrow$)  &  5.06  &  2.67    & 2.25  & 1.56 & 0.80 & $\textbf{0.50}$  \\
    SECS ($\uparrow$)  &  0.85  & 0.87    & 0.89  & 0.90 & 0.91 & $\textbf{0.92}$  \\
 \bottomrule
  \end{tabular}
  }\vspace{-0.6cm}
\end{table}
\subsection{One-shot VST}
\vspace{-0.1cm}We compared the performance with zero-shot and one-shot VST with different numbers of fine-tuning steps. Table \ref{oneshotVST} demonstrated that fine-tuning with one sample can improve the VST performance in terms of EER and SECS. However, the linguistic consistency decreased after overfitting to the small training samples so we only fine-tune the model with 1,000 steps.   
\begin{table}[t]
  \centering
  \caption{Results of ablation study on zero-shot VST scenario}\vspace{-0.3cm}
\label{ablationT}
      \resizebox{1\columnwidth}{!}{
  \begin{tabular}{l|c|cc|cc}
    \toprule
     Method & $p_{uncond}$ & CER   & WER & EER  &SECS    \\
    \midrule
HierVST (Ours) & 0 &  2.56& 5.86 & 6.73 & 0.843  \\
                & 0.1 &   2.12& 4.95 & \textbf{6.25} & \textbf{0.847}  \\
                & 0.2 &  2.04& 4.79& 7.77&0.838 \\ 
                & 0.5&  1.69&4.13& 8.25&0.836 \\
     \midrule
     $-$ PD &0 & 5.48& 11.81&8.5&0.835\\
     $-$ PD $-$ HVAE & 0 &  1.05&3.66& 11.75&0.816 \\ 
     $-$ PD $-$ HVAE $-$ HAG & 0 & \textbf{0.78}&\textbf{3.09}& 13.75&0.816 \\
    \bottomrule
  \end{tabular}
     }\vspace{-0.5cm}

\end{table}

\vspace{-0.1cm}\subsection{Ablation study}
\vspace{-0.1cm}\subsubsection{Hierarchical VAE}
\vspace{-0.1cm}We adopt the hierarchical VAE (HVAE) to restore the perturbed linguistic representation and to increase the speaker adaptation quality. Table \ref{ablationT} shows that removing the HVAE significantly decreases the performance of speaker adaptation. However, we found that the hierarchical structure requires more training steps to achieve the lower CER and WER for proper pronunciation in that the HierVST trained with 600k steps has a lower CER and WER. Also, the model has better naturalness which means that the hierarchical structure reduces the degradation of audio quality by regularizing an acoustic representation with a speaker-related linguistic representation. Note that it is necessary to perturb the waveform audio to remove the speaker-relevant information in the linguistic representation, so the model trained without audio perturbation is not able to convert the voice style.       
\subsubsection{Hierarchical adaptive generator}
\vspace{-0.1cm}We modify the HiFi-GAN by combining it with the source generator. With the distillation of the source-related representation, the model with a HAG synthesizes audio with better quality as indicated in Table \ref{ablationT} and the adaptation performance also increased regarding EER.

\vspace{-0.1cm}\subsubsection{Prosody distillation}
\vspace{-0.1cm}Although hierarchical VAE can improve the VST quality, the model has a higher CER and WER. Therefore, we additionally introduce prosody distillation (PD) for enhanced linguistic representation. Adding prosody distillation improves the overall performance regarding all metrics with an enhanced linguistic representation. We also compared the 20 bins of Mel-spectrogram with the full-band of the Mel-spectrogram, and the model trained with 20 bins of Mel-spectrogram has a lower F0 $l1$ distance in the source generator, therefore, we used only 20 bins for the prosody distillation.

\vspace{-0.1cm}\subsubsection{Unconditional generation}
\vspace{-0.1cm}We train the model with unconditional generation on the HAG with different unconditional ratios. We found that increasing the unconditional ratio improved the pronunciation of the converted speech. However, a model with a small ratio could generate converted speech with better speaker adaptation. Table \ref{ablationT} shows that adopting an unconditional generation with a proper ratio simply improved the model capacity for generation tasks.

\vspace{-0.1cm}\section{Conclusion}
\vspace{-0.1cm}We present HierVST, which can convert speech by hierarchically transferring the voice style. With only a speech dataset, we restored the linguistic representation from the disentangled representation, reproduced the enhanced linguistic and rich acoustic representation, and generated high-quality converted speech. Furthermore, we improve the capacity of the entire model using prosody distillation and unconditional generation. The experimental results demonstrated that our model can generate converted speech with high-fidelity audio and high-quality speaker adaptation. We see that our hierarchical adaptive structure can be adopted in unit-based speech-to-speech translation systems to generate an expressive voice style of translated speech. Although our model generates high-quality converted speech, our model has little controllability without converting the timbre. In future works, we will utilize pitch and duration to directly control the intonation and rhythm of speech.            

\vspace{-0.1cm}\section{Acknowledgements}
\vspace{-0.1cm}This work was supported by Institute of Information \& Communications Technology Planning \& Evaluation (IITP) grant funded by the Korea government (MSIT) (No. 2019-0-00079, Artificial Intelligence Graduate School Program (Korea University) and No. 2021-0-02068, Artificial Intelligence Innovation Hub) and Clova Voice, NAVER Corp., Seongnam,
Korea.

\bibliographystyle{IEEEtran}
\bibliography{mybib}

\end{document}